\documentclass[runningheads,a4paper]{llncs}

\usepackage{enumitem}
\setlist{nolistsep}

\usepackage{amssymb}
\usepackage{graphicx}
\usepackage[usenames,dvipsnames]{color}

\usepackage{url}

\usepackage{amsmath}
\usepackage{textcomp} 
\usepackage{xspace} 
\usepackage{bbding} 
\usepackage{pifont} 
\usepackage{stmaryrd} 
\usepackage{mdwlist} 
\usepackage[cp1252]{inputenc}

\begin{document}

\title{Applied Logic in Engineering 
} 
\author{Maria Spichkova}
\institute{RMIT University,  Australia \\  \email{maria.spichkova@rmit.edu.au} }
\maketitle

\begin{abstract} 
 Logic not only helps to solve complicated and safety-critical problems, but also disciplines the mind and helps to develop abstract thinking, which is very important for any area of Engineering.
In this technical report, we present an  overview of common challenges in teaching of formal methods and discuss our experiences from the course \emph{Applied Logic in Engineering}. This course was taught at TU Munich, Germany, in  Winter Semester 2012/2013.
\end{abstract}

\section{Introduction}

David Parnas, a pioneer of Software Engineering, stated that a solid understanding of logic should be essential for a software engineer, cf. \cite{Parnas93}:
\emph{``Professional engineers can often be distinguished from other designers
by the engineers  ability to use mathematical models to describe and
analyze their products."}. 
This statement is also true for any kind of engineering, from Software to Mechanical Engineering. However, (i) this subject is in most cases avoided by students as ``too boring" and ``too complicated" as well as (ii) the formal methods are often treated by engineers as ``something that is theoretically important but practically too hard to understand and to use", where the second is in many aspects a consequence of the first.

Unfortunately, dealing with formal methods often assumes that only two factors must be satisfied: the method must be sound and give such a representation, which is concise and beautiful just from the mathematical point of view, without taking into account any question of readability, usability, or tool support. This leads to the fact that the term ``formal" is for many people just some kind of synonym for ``unreadable", however, even small syntactical changes of a formal method can make it more understandable and usable for an average engineer. In our work on \emph{Human Factors of Formal Methods} \cite{hffm_spichkova,spichkova2013design} 
 we aim to apply the engineering psychology achievements to the design of formal methods, focusing on the specification phase of a system development process.
 
Dealing with  Formal Methods require a mathematical background and abstract thinking skills, which brings us to another problem: 
many students have negative perceptions and even fear of courses that require dealing with complex mathematical notations. This is strongly related to the phenomenon of \emph{mathematical anxiety}, cf. \cite{MathsAnx_Wang,sherman2003mathematics}.  The term  \emph{mathematical anxiety} was introduced in 1972 by Richardson and Suinn as
\emph{``feelings of tension and anxiety that interfere with the manipulation of numbers
and the solving of mathematical problems in a wide variety of ordinary life and
academic situations,"} cf. \cite{richardson1972mathematics}.
As stressed by Wang et al.,  mathematical anxiety has attracted
recent attention because of its damaging psychological effects and potential associations with mathematical problem
solving and achievement. 

Thus, the usability improvement is only a partial solution to the problem. To overcome the preconceived notions about formal methods we should start a bit earlier as on the development stage, by trainings and teaching of logic not only by presenting its theoretical aspects but also focusing on its real applications, industrial and non-industrial ones, referring to the programming languages where the formal side is almost covered, or to famous fiction books and movies, e.g., to the famous  crime stories by A.C. Doyle.

A novel way to attract students while teaching Formal Methods was presented in  \cite{curzon2013teaching}.
 Within the engagement project \emph{cs4fn}, Computer Science for Fun,   
the authors taught logic and computing concepts using magic tricks, which inspired students to work with logical tasks. 
Our approach was less revolutionary: we based the course on both practical examples and entertainment examples, such as formal modelling of Sherlock Holmes deductions.

\section{Applied logic in Engineering}

The course \emph{Applied logic in Engineering} (ACE) was introduced at TU Munich, Germany, in Winter Semester 2012/2013 as a face-to-face course.\footnote{\url{http://www4.in.tum.de/lehre/vorlesungen/Logic/WS1213/index.shtml}}
 
In contrary to the many courses on Formal Methods, we did not expect any previous knowledge on logics and abstract thinking. 
We introduced this lecture course as a ``logic for everybody", to engage pupils studying an engineering subject and is interested in getting the knowledge about logic and its application areas. However, this course would be especially beneficial for Computer Science students, as well as for the IT students who aim to work as Requirements Engineers and Testers. 
 
The course  
has the following \emph{learning outcomes}: On completion of this course students (1) will be able to state the basic principles of logic applied in Engineering  and (2) will experience practical applications of these principles.

To explain the core ideas of basic logics, we provided illustrative examples. 
Some of the examples will come from industrial problems, where some of the examples 
will have more entertainment nature (to show to the students that logic is not necessary ``a very dry subject") be presented by puzzles and analysis of situations from famous fiction books and movies.

\begin{figure}[ht!]
\begin{center}
\includegraphics[scale=0.4]{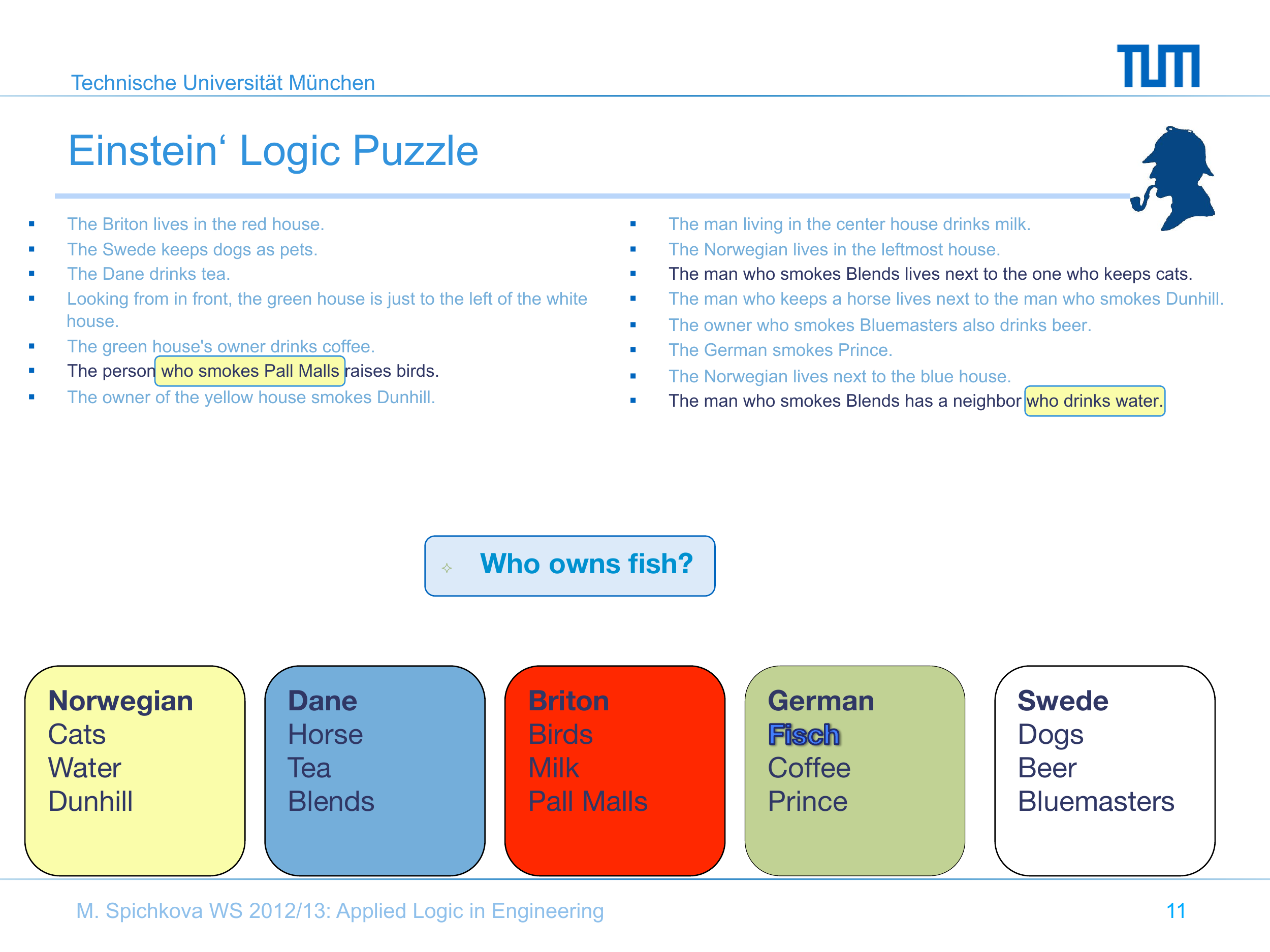}
\end{center}
\vspace{-5mm}
\caption{Solving the Einstein puzzle: A task for formal methods}
\label{fig:Einstein}
\end{figure}

\begin{figure}[ht!]
\begin{center}
\vspace{10mm}
\includegraphics[scale=0.4]{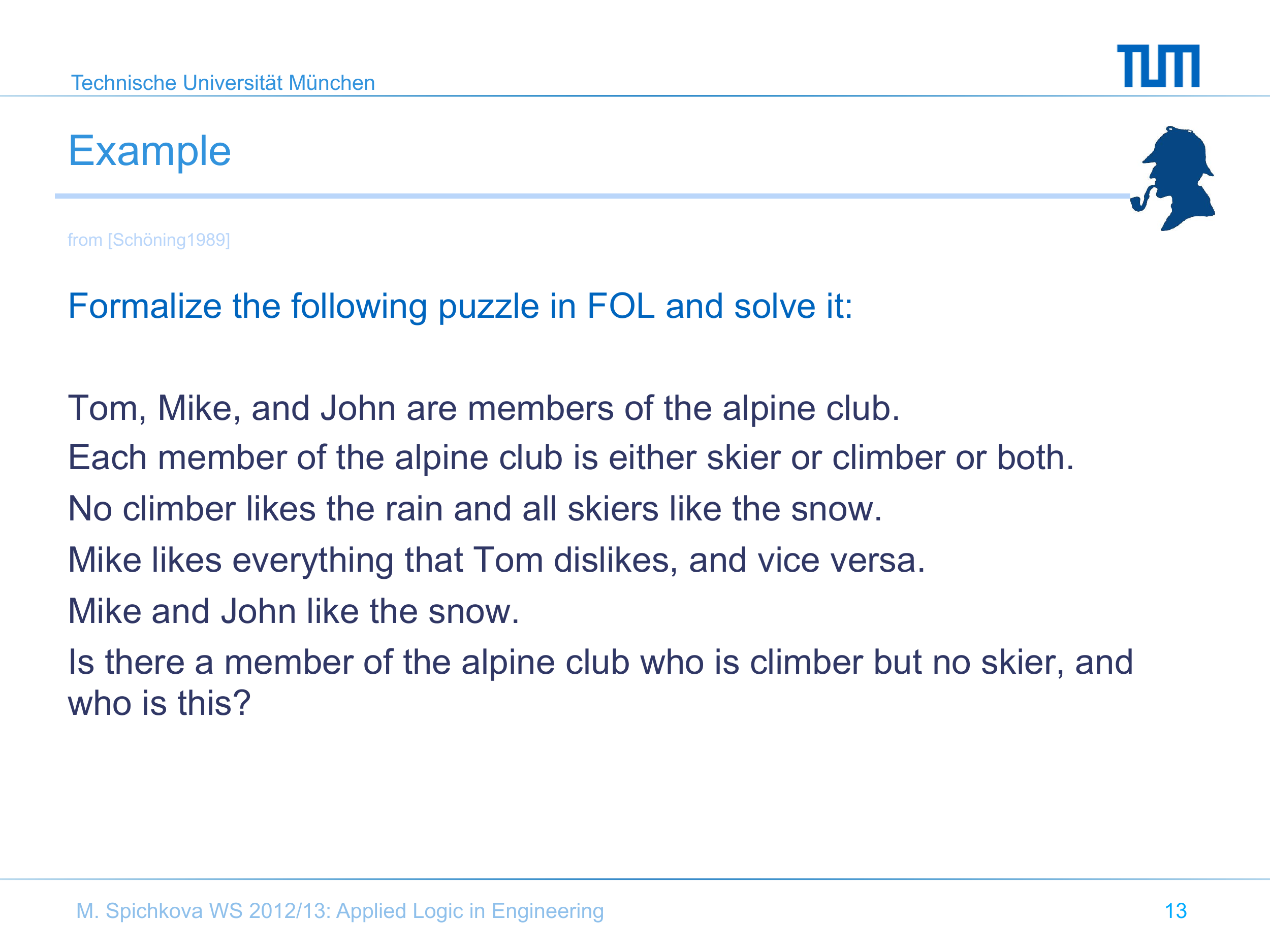}
\end{center}
\vspace{-5mm}
\caption{Solving a puzzle in First-Order Logic (FOL)}
\label{fig:puzzle}
\end{figure}

The course was partially based on the book of Sch{\"{o}}ning \cite{Logic4CS} ,  which introduces the notions and methods of formal logic from a computer science standpoint, as well as on the book of  Russell and Norvig  \cite{AIbook}.
 We also recommended our students to read the book of  Harrison, which focuses on practical application of logic and automated reasoning \cite{practLogic}, as well as a number of other books on logic and (semi-)automated theorem proving, cf.  \cite{LogicCS,FOL_ATP,aussagenlogik}.

\newpage
\noindent
The general structure of the course is presented below:\\
\begin{itemize}
\item{Part 1.}
\emph{What is Logic?}
	\begin{itemize}
	\item History and application areas
	\item ``Dry formal methods" vs. ``Sherlock Holmes deduction methods"
	\item Abstract representation and abstract thinking
	\item ProLog 
		\begin{itemize}
		\item Another way of thinking (in comparison with C, Java, etc.)
		\item Current projects (e.g., at Siemens AG)
		\end{itemize}
	\item Overview of logics: propositional, first order/predicate, fuzzy, Higher-Order Logic
	\end{itemize}
~%
\item{Part 2.}
\emph{Propositional logic}
	\begin{itemize}
	\item  Syntax and semantics
	\item Normal form transformation
	\item Calculi 
		\begin{itemize}
		\item Natural deduction 
		\item Resolution
		\end{itemize}
	\item Binary decision diagrams
	\item ProLog representation
	\end{itemize}
~%
\item{Part 3.}
\emph{First Order/ Predicate logic}
	\begin{itemize}
	\item Syntax and semantics
	\item Normal form transformation
	\item Substitution
	\item Calculi
		\begin{itemize}
		\item Natural deduction  
		\item Resolution
		\end{itemize}
	\item ProLog representation
	\end{itemize}
~%
\item{Part 4.}
\emph{Special Cases}
	\begin{itemize}
	\item Datalogic and databases 
	\item Description Logic and Entity-Relation Diagram 
	\item Closed world assumption
	\end{itemize}
~%
\item{Part 5.}
\emph{Applications}
	\begin{itemize}
	\item Formal specification and verification
		\begin{itemize}
		\item Algebraic specification 
		\item Specification in Attempto Controlled English  
		\item HOL specification (Microsoft: Hypervisor Verification in VerisoftXT)
		\end{itemize}
	\item Reasoning, problem of planning
		\begin{itemize}
		\item Event calculi
		\item FLUX language (ProLog + current projects)
		\end{itemize}
	\end{itemize}
\end{itemize}

\newpage

\begin{figure}[ht!]
\begin{center}
\includegraphics[scale=0.45]{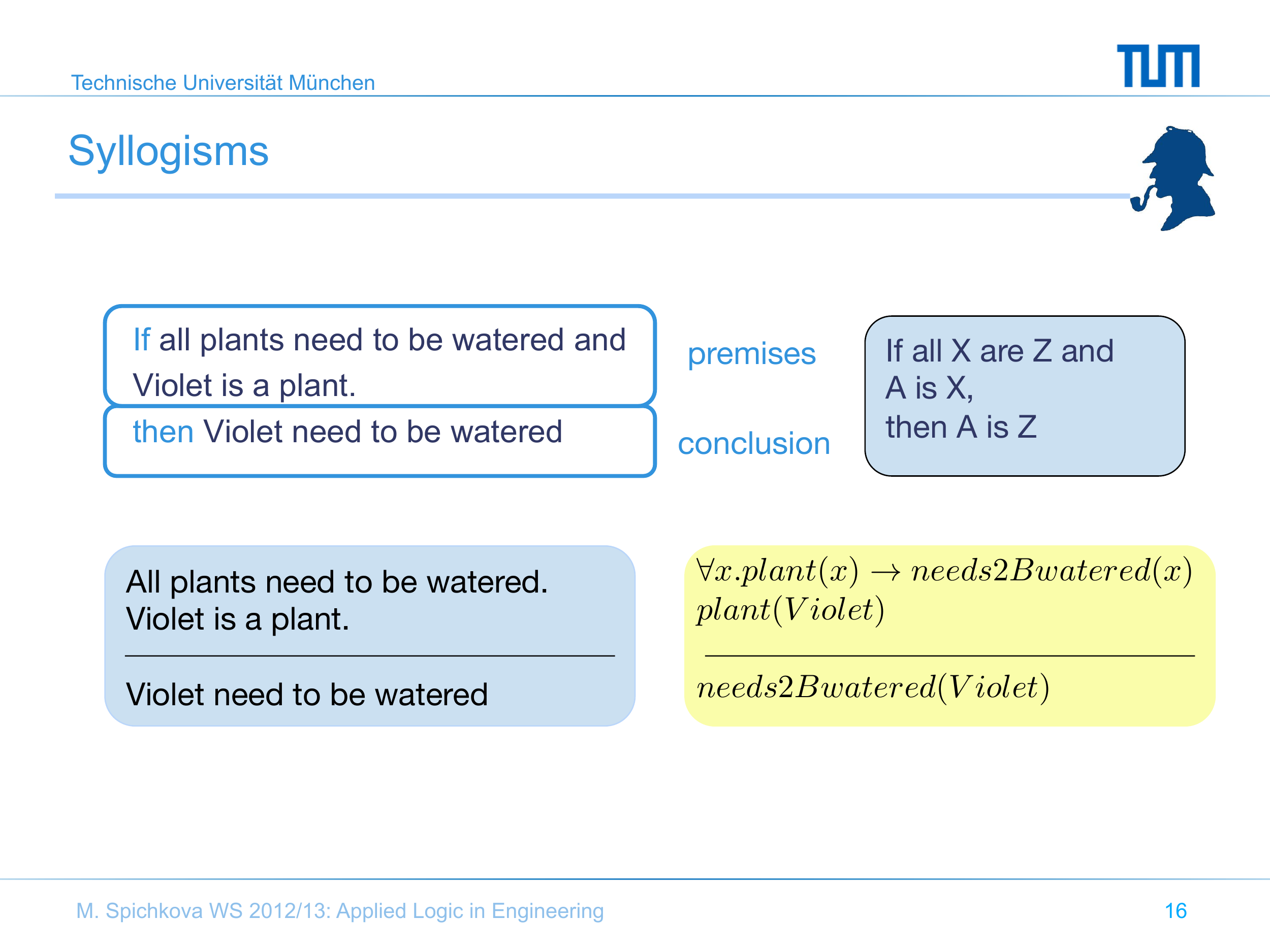}
\end{center}
\vspace{-5mm}
\caption{Visual explanation of formal notation: Introduction to the Syllogisms}
\label{fig:Syllogisms}
\end{figure}

~\\~\\
\noindent
Thus, the course gives an introduction not only to the basic principles of logic, but also to some of its applications, such as 
\begin{itemize}
\item Reasoning and Planning problems;
\item Formal Specifications/ models for precise description of systems and requirements
and analysis of systems;
\item Verification: Proving that a system fulfils its requirements, and that 
 a new version of a system is a refinement of the previous version;
\item Theorem proving/Model checking allowing 
  (semi-)automated proofs;
\item  Design/optimization of digital circuits:
Claude Shannon  has shown that propositional logic can be used to
describe and optimize electromechanical circuits, \cite{Shannon1937};
 \item  Formalisation of queries in databases.
\end{itemize}
We also discussed application of formal methods in a number of recent research projects, cf. \cite{spichkova2012verified,botaschanjan2008correctness,spichkova2006flexray,kuhnel2006upcoming,kuhnel2006flexray,feilkas2011refined,holzl2010autofocus,feilkas2009top}.

\newpage
\noindent
\textbf{Example.}
Let us discuss an example of a task for tutorial:\\
\emph{
	Formalize the following sentences as formulas and then show that 
	they are equivalent:
	\begin{itemize}
		\item[(1)] The following property holds not for all time intervals: 
		If the system gets a signal from its sensors that there is no communication  at a time interval $t$  or that  the 
					battery power gets low  at a time interval $t$, and exists an information package that have to be send, then at a time interval $t$
					there is an information package in the temporal buffer.
		\item[(2)] At some time interval $t$ the following holds for all information packages:
		there is an information package that have to be send, but
		there is no information package in the temporal buffer, and 
		the system gets a signal from its sensors that there is no communication   or that  the 
					battery power gets low.
	\end{itemize}
	}
\noindent
~\\
One possible solution:\\
Formalisation of the sentences would be	\\
(1) $\neg\forall t. \left( (C(t) \vee B(t)) \wedge S(t) \to T(t) \right)$  and\\
(2) $\exists t. \left(  S(t) \wedge \neg T(t) \wedge (C(t) \vee B(t))  \right)$.\\
Proof that both formulas are equal:
\\
$\neg\forall t. \left( (C(t) \vee B(t)) \wedge S(t) \to T(t) \right)$
\\
$\equiv \exists t. \neg \left( (C(t) \vee B(t)) \wedge S(t) \to T(t) \right)$
\\
$\equiv \exists t. \neg \left( \neg ( (C(t) \vee B(t)) \wedge S(t))  \vee T(t) \right)$
\\
$\equiv \exists t. \left(  ( (C(t) \vee B(t)) \wedge S(t))  \wedge \neg T(t) \right)$
\\
$\equiv \exists t. \left( S(t) \wedge \neg T(t) \wedge  (C(t) \vee B(t))  \right)$
\\[5mm]
Another possible solution:
\\
Formalization of (1): $\neg\forall t. \exists p. \left(  (C(t) \vee B(t)) \wedge S(p,t) \to T(p,t)  \right) $
\\
Formalization of (2): $\exists t. \forall p. \left(  S(p,t) \wedge \neg T(p,t) \wedge (C(t) \vee B(t))  \right)$
\\
Proof that both formulas are equal:
\\
$\exists t. \forall p. \left(  S(p,t) \wedge \neg T(p,t) \wedge (C(t) \vee B(t))  \right)$
\\
$\equiv \neg \forall t. \neg (\forall p. \left(  S(p,t) \wedge \neg T(p,t) \wedge (C(t) \vee B(t))  \right))$
\\
$\equiv \neg \forall t. (\exists p. \neg \left(  S(p,t) \wedge \neg T(p,t) \wedge (C(t) \vee B(t))  \right))$
\\
$\equiv \neg \forall t. (\exists p. \left(  \neg S(p,t) \vee T(p,t) \vee \neg (C(t) \vee B(t))  \right))$
\\
$\equiv \neg \forall t. (\exists p. \left(  \neg S(p,t) \vee \neg (C(t) \vee B(t)) \vee T(p,t)  \right))$
\\
$\equiv \neg \forall t. (\exists p. \left(  \neg (S(p,t) \wedge (C(t) \vee B(t))) \vee T(p,t)  \right))$
\\
$\equiv \neg \forall t. (\exists p. \left(  (S(p,t) \wedge (C(t) \vee B(t))) \to T(p,t)  \right))$

\hspace{111mm}$\Box$

~\\
ACE was introduced as an elective course and attracted 20 students.
The exam for this course was organised as an \emph{open book} exam, as our goal was to examine whether the students understand and are able to apply the core principles of logic methods, rather than check they memory.

\section{Evaluation and  Conclusions}

As per evaluation report \cite{ace201213}, the majority of the students agreed that the provided examples were very helpful, and the learning amount and the amount of the material provided within the course  were ``exactly right" (German, ``genau richting").
We got the following comments from our students:
\\
\emph{``Structured logically and builds up stuff part by part; nice additions as Sherlock video"};
\\
\emph{``The topic presented are interesting and indeed ``applied", unlike other logical courses that are more theoretic"};
\\
\emph{``I liked the small size of the douse and I got a deeper understanding of logic"}.
To the question what did you most liked in the course, the students replied\\
\emph{``Sherlock, Examples during lecture"} . 

This technical report present an  overview of common challenges in teaching of formal methods and suggested solutions to them.
We discussed our experiences from the course \emph{Applied Logic in Engineering} taught at TU Munich, Germany.
  
\bibliographystyle{abbrv}

\end{document}